# VIQI: A New Approach for Visual Interpretation of Deep Web Query Interfaces

Radhouane Boughammoura
Department of Computer Sciences
Faculty of Sciences of Monastir,
Research Unit MARS
Monastir, Tunisia
Radhouane.Boughammoura@gmail.com

Lobna Hlaoua
Department of Electronic and Computer Science
High School of Sciences and Technology of H.Sousse,
Research Unit MARS
H. Sousse, Tunisia
Lobna1511@yahoo.fr

Mohamed Nazih Omri
Department of Computer Sciences
Faculty of Sciences of Monastir
Research Unit MARS
Monastir, Tunisia
MohamedNazih.Omri@fsm.rnu.tn

*Abstract*— **Deep Web databases contain more than 90% of pertinent information of the Web. Despite their importance, users don't profit of this treasury. Many deep web services are offering competitive services in term of prices, quality of service, and facilities. As the number of services is growing rapidly, users have difficulty to ask many web services in the same time. In this paper, we imagine a system where users have the possibility to formulate one query using one query interface and then the system translates query to the rest of query interfaces. However, interfaces are created by designers in order to be interpreted visually by users, machines can not interpret query from a given interface. We propose a new approach which emulates capacity of interpretation of users and extracts query from deep web query interfaces. Our approach has proved good performances on two standard datasets.**

## I. INTRODUCTION

Deep Web is the part of web hidden behind query interfaces. Deep Web is growing and changing continuously, new web services are emerging at a high rate, they offer new and complementary services to existing ones. For commercial reason, web services are increasingly competitive. Hence, users are interested by asking services of many providers at the same time. However, web services have different query capabilities, users have to formulate a query for each web service separately. Moreover, as each web service query has its own meaning, results will be incoherent.

Classical Information Retrieval process has 4 steps. For example a student attending NCM 2012 conference in Seoul, Korea will be interested by flights to Seoul and will proceed as follows(seefigure1):

This work is supported by research unit MARS, FSM

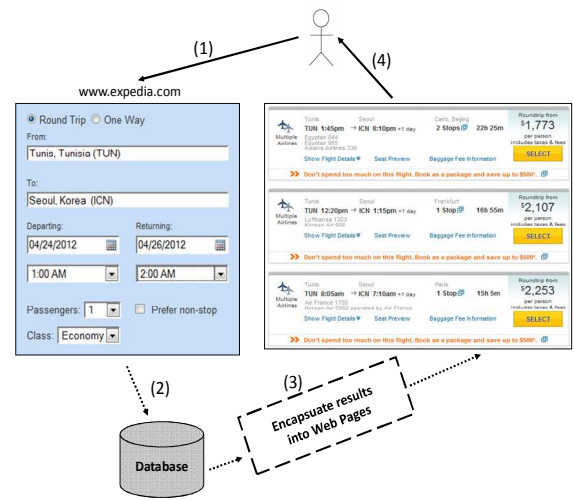

Figure 1. Classical Information Retrieval Process from deep Web

(1) he formulates a query using query interface, (2) the query is submitted to deep Web database, (3) relevant information are extracted from database and encapsulated into web pages, (4) finally web pages are returned to users.

We believe it is possible to make web services more sensitive to user's needs. The system that we are working on collects querying capabilities of many web services and combines them in one query interface. Using this interface, user can formulate one query and obtains results satisfying his need from all web services. This interface is generic as it meets together many services and respects the autonomy of each service.

Figure 2 show on the left classic information retrieval process where user search information from each web service separately. On the right, the generic web service [1,2] where user formulates only one query, the system translates the query [3,4,5] to each local service and collects relevant results [16, 17] from all web services at the same time.

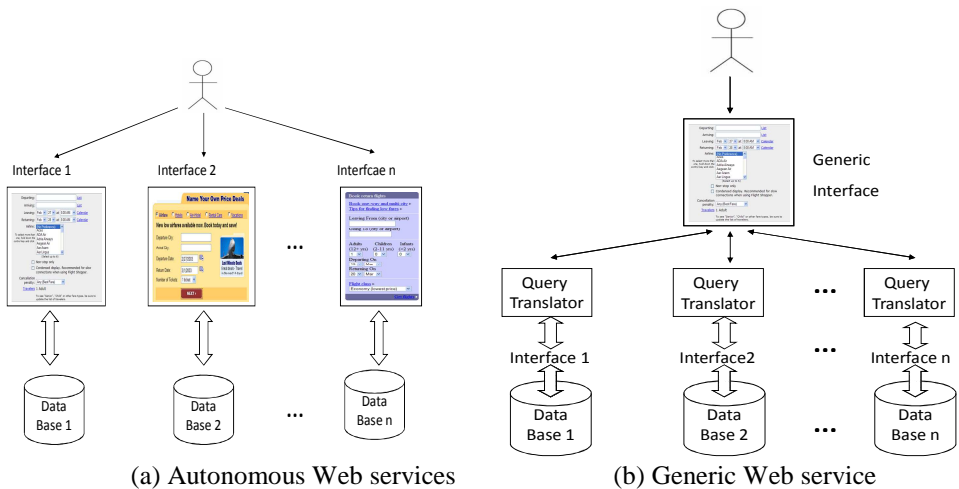

(a) Autonomous Web services      (b) Generic Web service

Figure 2. Overview of (a) autonomous web services and (b) unified web services

Query interfaces are the natural representation of the query for visual user perception [6, 7]. Although query interfaces are easily interpreted by users, they are not the query supported server's side. Query interfaces are used to map the query to URL which contains a list of attribute/value pairs. For example 'http://www.expedia.com/FlightsSearch?trip=''roundtrip''&leg1=''from:TUN,to:ICN'',departure=''04/24/2012''...' is query which aims to find flights from departing city to destination city. Attribute/value pairs are listed as a sequence: trip=''roundtrip'', leg1=''from:TUN,to:ICN'', departure = ''04/24/2012'', etc. URL is the form of the query which is run by web server. However, users have great difficulty to interpret and to understand the meaning of such query because fields are represented by internal names which are concatenated and abbreviated. So, although query interface have rich semantic value, web server can not run it, while URL have poor semantic meaning, but it can be run by web server.

Then to resolve this challenge, we propose three main contributions:
1. A new model for query representation: this model provides matching between elements of query and elements of query interface.
2. A new approach of query interpretation and extraction: our approach emulates capacity of users to interpret query interfaces
3. To evaluate our approach on two standard datasets

Our paper is structured as follows. Section 2 is related works. We present a new model for query representation detailed in section 3, which provides matching between interface and query elements. In Section 4, we describe our approach to extract query from a given query interface based on our model of query representation. We evaluate performance of our approach in section 5. Finally we conclude and present our future work in section 6.

II. RELATED WORK

Related works have used two main query representation models: a flat model and a hierarchical model.

According Madhavan et al [8], flat representation of query is represented by sequence of attribute/value pairs. This category of queries cannot be interpreted visually by user. Hence, users have difficulty to understand the meaning of query. However, they are indexed by standard search engines such as Google [9] as static resources.

While flat queries are not understood by users, hierarchical queries [10,11] are easily interpreted (see figure 2), they describe visual structure of query interface, i.e. all concepts of the query and semantic relations between them. As hierarchical query is not the query but its structure, they are not indexed by standard search engines. Despite their rich semantic values, hierarchical queries still unknown.

Approaches of query extraction can be classified in two categories: Approaches based on visual features [12,13,14], and approaches based on HTML features [15,16,17].

Visual features include topological relations between fields, direction, distance, etc. They are detectors of semantic concepts of query. For example, there is a high probability that two adjacent (distance and direction) fields form the attributes of the same concept.

While two fields are adjacent in web page space, they are too far from each other in HTML script because of HTML tags, field's values, etc. HTML based approaches use linguistic features to extract concepts of the query. For example, 'departure' and 'destination' are attributes of the same concept because they are instance of the same linguistic concept 'City'.

III. MODEL OF REPRESENTATION OF QUERY

We have chosen to represent query by a hierarchical model [18] as this model reflects meaning of query (see figure 3). Concepts of query are rendered in query interface based on a spatial-locality paradigm: fields which form attributes of the same concept are close to each other in the interface and aligned (e.g. 'Adults' and 'Children'), while fields of different concepts are rarely adjacent and aligned (e.g. 'Adults' and 'ReturningMonth').

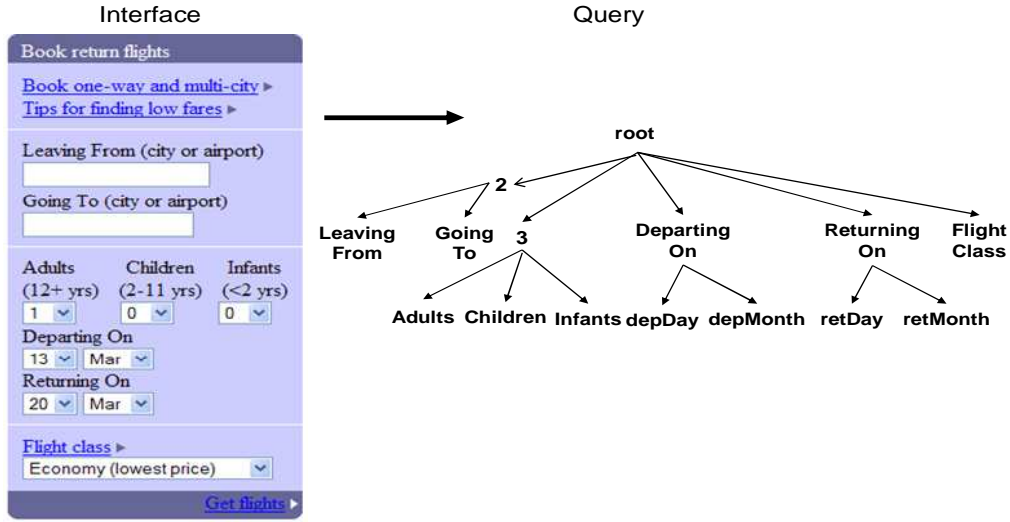

**Figure 3. Hierarchical query extracted from query interface (Left)**

Based on mapping between visual features and semantic concepts of the query (see figure 4), we have found the five components of our model: *field*, *RenderedGroup*, *RenderedCollection*, *NotRenderedGroup*, and *VisualBox*:

- *field*: this is the basic unit of information composing the query, it is a query condition aver one attribute of the query. This component is rendered as a rectangular box in web page space where user can give some input information
- *RenderedGroup*: it represents one concept of the query, it conatins a list of attributes. Each attribute may be recursively another RenderedGroup or a field. It is rendered in web page as recursive imbrications of rectangular boxes.
- *RenderdCollection*: it is the root of the query, it meets all concepts of the query. It is rendered in web page as the most external rectangular box.
- *NotRenderedGroup*: Some elements in web page such as pictures and hyperlinks are not in the query. There is no mapping between these elements and attributes of the query.
- *VisualBox*: one internal element of the query may contains concepts of different natures (field, group of fields, super-group). Hence, in order to be grouped together, all elements of the model extend one abstract visual element: the VisualBox.

*Example:* Query in figure 2 is instance of our query model. There is ten fields, four RenderedGroup (2, 3, Departing On, Returning On), and one RenderedCollection (root).

IV. A NEW APPROACH OF QUERY EXTRACTION

The basic idea of VIQI (Visual Interpretation of Query Interfaces) is based on *User's Interpretation*. We call User's Interpretation the visual human capacity to distinguish group of fields in web page space. Interpretation is closely related to proximity and alignment between fields. For example, the group of fields {''Leaving From'', ''Going To''} forms one concept of the query and are close to each other and left aligned (see figure 3).

We measure closeness between fields by Euclidean Distance between them. The Euclidean Distance between two fields is the minimal distance between any pair of points which belong respectively to each one of the fields (see Equation 1).

$$DistEucl(f1, f2) = \min_{p1 \in f1, p2 \in f2} DistEucl(p1, p2) \quad (1)$$

with p1 a point of f1 and p2 a point of f2

We calculate alignment between fields based on *Align* function. We distinguish four classes of alignments: Buttom, Top, Left, and Right. The function *AlignX* returns 1 when two fields are aligned, else it returns 0.

$$AlignX(f1, f2) = \begin{cases} 1 & \text{if } f1 \text{ et } f2 \text{ are aligned X} \\ 0 & \text{else} \end{cases} \quad (3)$$

X takes the values 'B' (Buttom), 'T' (Top), 'L' (Left), or 'R' (Right).

$$Align(f1, f2) = 2.AlignB(f1, f2) + AlignT(f1, f2) + AlignL(f1, f2) + AlignR(f1, f2) \quad (2)$$

Over the alignment of fields is high, fields are more likely to belong to the same concept. Buttom alignment is the most descriptive to grouping of fields because reading order of fields is per line (buttom alignment). Equation 4 gives the measure Proximity. This measure depends on two factors: Euclidian distance and Alignment. Over closeness of fields

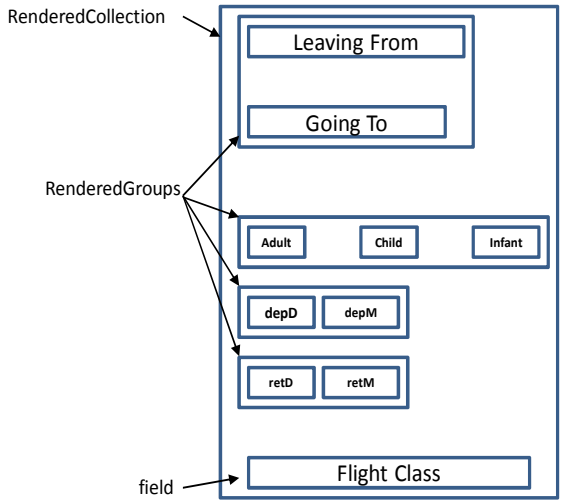 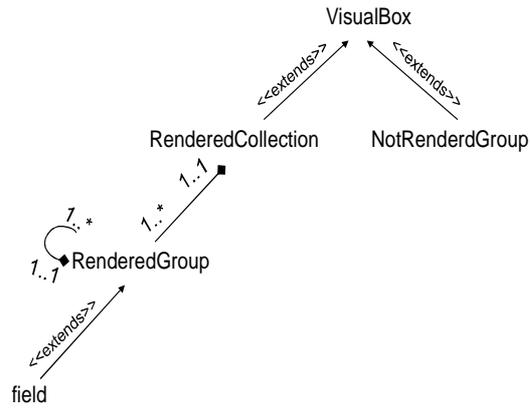

**(a) Visual Boxes of Query**  **(b) Query Model**

**Figure4. Matching visual query elements to query's semantic concepts**

is high, the Euclidean distance between them is small, then the proximity between fields is also small. Hence fields have high probability to be grouped together. In other hand, over alignment of fields is high, Align (f1, f2) is High, then the proximity between fields is small. Hence probability to be grouped together is high.

$$\text{Proximity}(f1, f2) = \frac{DistEucl(f1, f2)}{Align(f1, f2)} \quad (4)$$

In order to calculate groups of fields, we have used a clustering algorithm based on density DBSCAN (Density-Based Spatial Clustering of Applications with Noise) [19]. The definition of cluster in DBSCAN is based on the concept of scope of density: a field f1 is in the scope of density of field f2 if the proximity between f1 and f2 is inferior to ε. We have fixed ε to the minimal proximity value between any two fields not clustered, and Noise is orphan fields, i.e it has no other fields in its scope of density (e.g field 'FlightClass'). DBSCAN algorithm is shown in Algorithm 1. In order to detect clusters, DBSCAN proceeds as follows:

- step1. Choose one field not clustered (Line 4)
- step2. Get its immediate neighbors (Line 5)
- step3. If field has no neighbors then mark it as Noise (Line 7), else expand the cluster recursively to the agglomeration of fields (Line 13)
- step4. Repeat step1, step 2, and step3 until all fields are clustered

Figure 4 is a running example of the algorithm on query interface of figure 2. Squares correspond to fields and circles show the scope of density of each field. One field is in the scope of density of a second field if its corresponding circle reaches the center of the second field's square.

```
0)DBSCAN(D, eps, MinPts)
1)Output: Cluster C
2)BEGIN
3)Cluster C_loc = Φ
4) for each unvisited field f in query
5)    N = scopeDensity(f, eps)
6)    if sizeof(N) < MinPts
7)       mark f as NOISE
8)    else
9)       Cretae cluster C_loc
10)      add f to C_loc
11)      mark f as visited
12)      for each unvisited field f' in N
13)         expandCluster(f', N, C_loc, eps, MinPts)
14)      ENDFOR
14)      add C_loc to C
15)   ENDELSE
16) ENDFOR
17)END

I) ExpandCluster(f', N, C_loc, eps, MinPts)
II)    BEGIN
III)      N' = scopeDensity (f', eps)
IV)       if sizeof(N') < MinPts
V)           mark f' as NOISE
VI)       else
VII)         mark f' as visited
VIII)        add f' to C_loc
IX)          for each unvisited field f'' in N'
X)              ExpandCluster(f'', N', eps, C_loc)
XI)          ENDFOR
XII)      ENDELSE
XIII)END
```

**Algorithm 1. Algorithme DBSCAN**

For example, 'Adults' scope of density reaches 'Children' field, and 'Children' scope of density reaches 'Adults' and 'Infants'. One iteration of DBSCAN creates the group of fields 'NumberPassengers' as follows:

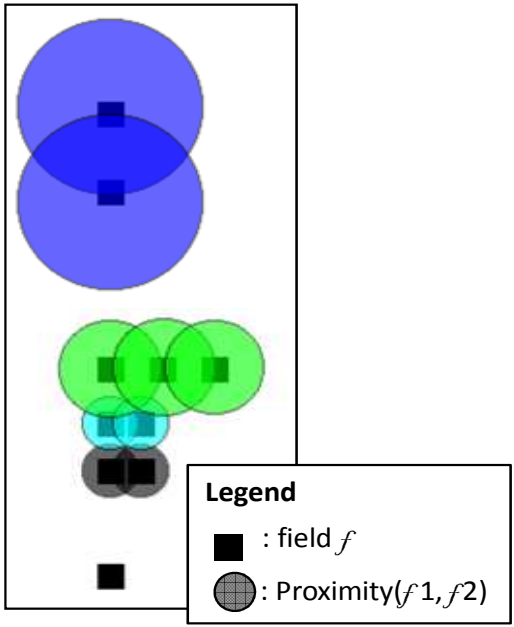

**Figure 4. Running Example of DBSCAN**

- In step1, field 'Adults' is chosen.

- In step2, scope of density of 'Adults' reaches 'Children' (see 'Adults' green circle), so the cluster is {'Adults', 'Children'}

- In step3, the cluster is expanded to {'Children', 'Infants'} because scope of density of 'Children' reaches 'Infants' (see 'Children' green circle), so the cluster {'Adults', 'Children'} is expanded to {'Adults', 'Children', 'Infants'}.

- In step4, choose next not clustered field

We remark that circles having the same color form one group of fields which correspond to one concept of the query.

We have shown in the running example how to calculate group of fields. Super-groups are recursively obtained by running recursively DBSCAN on groups. Stop condition is reached when all fields are clustered into one super-cluster 'root'.

## V. EXPERIMENTAL RESULTS

We have tested performance of VIQI on two standard datasets ICQ and TEL-8 [20]. ICQ and TEL-8 are two collection of query interfaces collected from deep Web services. For each query interface, its manually extracted query is available on the dataset. Interfaces are classified into five classes of subjects: Airfare, automobile, Books, Real estate, and Jobs.

Our evaluation methodology is as follows: we extract query from query interface using VIQI, then we compare it to query extracted manually. If the two queries are the same, we say that VIQI have extracted the query correctly, else we say that VIQI have committed a mistake. Table 1 resumes our experimental results.

**Table 1. Experimental results**

|                | TEL-8 | | | ICQ | | |
|---|---|---|---|---|---|---|
|                | Airfare | Auto. | Books | Airfare | Auto. | Books |
| #interfaces    | 20 | 19 | 19 | 20 | 20 | 19 |
| #fields        | 10.75 | 7.78 | 5.35 | 10.70 | 5.10 | 5.30 |
| #correct query | 13 | 14 | 17 | 13 | 13 | 16 |
| #mistakes      | 7 | 5 | 2 | 7 | 7 | 3 |
| Precision      | 0,65 | 0,73 | 0,89 | 0,65 | 0,65 | 0,84 |

We remark that the precision of VIQI depends on the average number of fields in each collection of interfaces. Over number of fields in query interface is high, the more the precision of our algorithm is low. It depends also on the complexity of query: the more the imbrications of concepts are high, the more the precision of VIQI is low. Hence queries of airfaire subject are the more complex. Queries in Auto collection are less complex than airfare collection and more complex than Books' queries. The latter are the most simple because many queries are flat, i.e. all fields are clustered in one cluster. For Books collection VIQI have the highest precision (~ 90% on TEL-8).

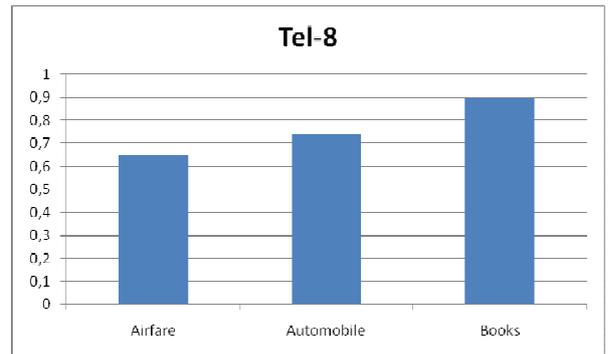

**Figure 2. Precision of VIQI on TEL-8**

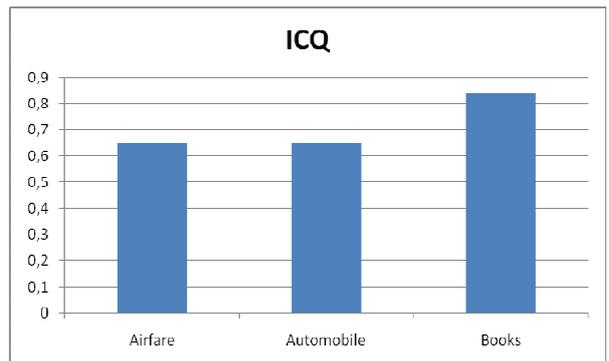

**Figure 3. Precision of VIQI on ICQ**

## VI. CONCLUSION

In this paper we have presented two main contributions. The first one is a new technique to represent concepts of query and semantic relation between them. The second contribution is a new approach of query extraction based on visual interpretation of interfaces, it extracts query from existing query interface. We have measured performance of our approach on two standard datasets ICQ and TEL-8 and we have proved that our approach achieves good precision.

Our future work is to propose an approach which integrates query interface of the same subject in one generic query interface. This approach is starting point to a new web service which facilitates Information Retrieval from Deep Web.

REFERENCES

[1] W. Wu, C. Yu, A. Doan, W. Meng, ''An interactive clustering-based approach to integrating source query interfaces on the deep Web'', In Proceedings of SIGMOD '04, 2004.

[2] W. Wu, A. Doan, C. Yu, ''Merging Interface Schemas on the Deep Web via Clustering Aggregation'', In *Proceedings of* ICDM '05, . 2005.

[3] F. Jiang, L. Jia, W. Meng, X. Meng, ''MrCoM: A Cost Model for Range Query Translation in Deep Web Data Integration'', In Proceedings of SKG '08, 2008.

[4] Z. Zhang, B. He, and K. C-C Chang, ''Light-weight domain-based form assistant: querying web databases on the fly'', In Proceedings of VLDB '05, 2005.

[5] Z. Zhang, B. He, and K. C.-C. Chang , ''On-the-fly Constraint Mapping across Web Query Interfaces''. In Proceedings of VLDB-IIWeb'04, 2004.

[6] J. Jansen and Dick C.A. Bulterman, ''Enabling adaptive time-based web applications with SMIL state'', In Proceedings of DocEng '08, 2008.

[7] M. Jayapandian, H. V. Jagadish. 2008, ''Expressive query specification through form customization'',  In Proceedings of EDBT '08, 2008.

[8] J. Madhavan, D. Ko, L. Kot, V. Ganapathy, A. Rasmussen, A.Y. Halevy, ''Google's Deep Web crawl'', In Proceedings of VLDB, 2008.

[9] www.google.com

[10] E-C. Dragut, T. Kabisch, C. Yu, U. Leser, ''A hierarchical approach to model web query interfaces for web source integration'', In Proceeding of. VLDB 2009, 2009.

[11] W. Wensheng, A-H Doan, C. Yu, W. Meng , ''Modeling and Extracting Deep-Web Query Interfaces'', In Proceedings of AIIS 2009, 2009.

[12] Z. Zhang, B. He, and K. Chang, ''Understanding Web query interfaces: Best-effort parsing with hidden syntax'', In Proceedings of SIGMOD'04, 2004.

[13] M. Dörk, S. Carpendale, C. Collins, C. Williamson, ''VisGets: Coordinated Visualizations for Web-based Information Exploration and Discovery'', IEEE Transactions on Visualization and Computer Graphics (Proceedings Information Visualization 2008), 2008.

[14] D. Cai, S. Yu, J-R. Wen, W-Y. Ma, ''Extracting content structure for web pages based on visual representation'', In Proceedings of the APWeb'03, 2003.

[15] H. Nguyen, T. Nguyen, J. Freire, ''Learning to extract form labels'', In  *Proceedings of. VLDB 2008, 2008*.

[16] R. Boughammoura, M-N. Omri, and H. Youssef., ''Fuzzy Approach for Pertinent Information Extraction from Web Resources'', Journal of Computing and e-Systems, Vol. 1 No. 1, 2008.

[17] R. Boughammoura, M-N. Omri, ''Statistical Approach for Information Extraction from Web Pages'', In proceedings of International Symposium on Distance Education (EAD'2009), 2009.

[18] R. Boughammoura, M-N. Omri, ''SeMQI: A New Model for Semantic Interpretation of Query Interfaces'', In Proceedings of  NGNS'11, 2011.

[19] M. Ester, H-P Kriegel, J. Sander, X. Xu, ''A density-based algorithm for discovering clusters in large spatial databases with noise'', In proceeding of KDD-96, 1996.

[20]  The UIUC Web Integration Repository, Computer Science Department, University of Illinois at Urbana-Champaign, http://metaquerier.cs.uiuc.edu/repository, 2003.